\documentclass[a4paper,11pt]{article}
\pdfoutput=1 

\usepackage{jinstpb} 
                     \usepackage[title]{appendix}

\usepackage{lineno}
\usepackage{xcolor}
\begin{document}

\title{\boldmath Spectrum of coherent VUV radiation generated by 5.7~MeV electrons in a multilayer X-ray mirror}

 \author[]{M.V.Shevelev,}
 \author[1]{S.R.Uglov,\note{ Corresponding author.}}
 \author[]{A.V.Vukolov}
 
 \affiliation{Tomsk Polytechnic University,\\Lenin Avenue 30, Tomsk 634050, Russia}

\emailAdd{uglov@tpu.ru}

\abstract{The spectral and angular properties of diffracted transition radiation (DTR) and parametric radiation (PXR) in the ultrasoft X-ray region generated by the periodic structure upon interaction with a relativistic electron beam with energy of 5.7 MeV are numerically studied using a sample of the periodic structure $[Mo/Si]^{50}$, known as a multilayer X-ray mirror. Based on calculations, an experimental approach is proposed to separate and identify the contributions of PXR and DTR. The ultrasoft X-ray radiation can be used to eliminate coherent effects occurring in the optical range when diagnosing the submicron electron beam size.}



\maketitle

\flushbottom

\section{Introduction}
{ Structures such as multilayer X-ray mirrors can be effective sources of vacuum ultraviolet and ultrasoft X-ray radiation with tunable radiation energy ~\cite{uglov6!,uglov12, uglov8andre}. The radiation band of the periodic structure (RPS) is determined by the period of the multilayer structure, the number of layers, and the tilt angle of the structure $\theta_0$ to the electron beam direction. The main contribution to the radiation is made by diffracted transition radiation (DTR) and coherent radiation of the periodic structure of the target excited by an electron beam – an analog to parametric X-ray radiation in crystals (PXR). A short-wave vacuum ultraviolet source with a quasi-monochromatic spectrum induced by the periodic structure of the target can also be used to develop methods for diagnosing and imaging submicron electron beams.  The use of shorter wavelength radiation will mitigate the impact of coherent effects characteristic of methods that employ optical radiation ~\cite{loos10a,wesch10b} and reduce the contribution to imaging of the fundamental diffraction limit ~\cite{kube10}. In this paper, we present the numerical simulations of the properties of the RPS consisting of PXR and DTR generated in the VUV region, based on the theoretical works ~\cite{nasonov5,nasonov6}. These studies were carried out using a periodic target consisting of 50 pairs of Mo/Si layers with a period d=11.32 nm interacting with the electron beam with a total electron energy of 5.7 MeV. The second section, based on model calculations as an example, examines the formation of angular distributions depending on DTR and PXR contributions, as well as the spectral properties of each of the contributions. The third section discusses the feasibility of experimental studying the fine structure of the spectrum and estimating the spectral line widths.

    \begin{figure}[htp]
    \centering
    \includegraphics[width=60mm]{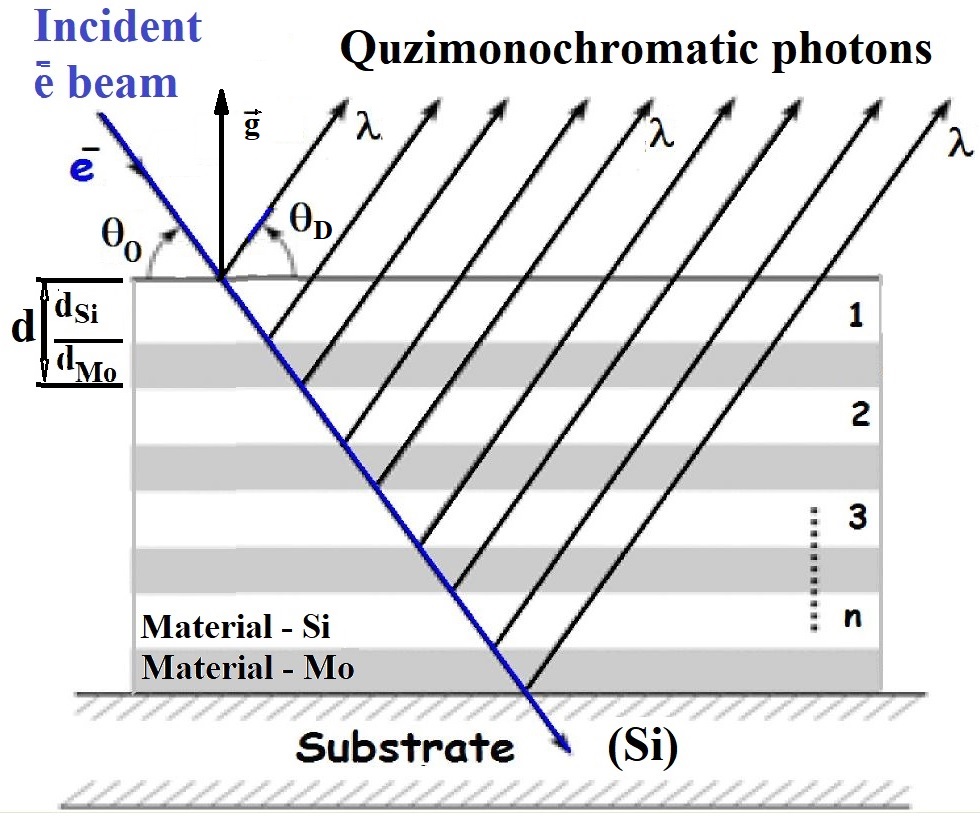}
    \caption{Scheme  of radiation generated by a multilayer periodic structure.}
    \label{fig:g1}
\end{figure}
}

\section{General properties of DTR and PXR contributions }

       One of the fundamental parameters of the periodic target that determines the properties of DTR and PXR is the periodic dielectric susceptibility.
\begin{equation}
     \chi(\hbar\omega, \vec r) =\chi_0(\hbar\omega) + \Sigma_{\vec g}\chi_g(\hbar\omega)e^{i \vec g \vec r}.
\end{equation}
     
  Here  $\chi_0(\hbar\omega) =(a \chi_a(\hbar\omega)+b\chi_b(\hbar\omega))/d$ is the dielectric susceptibility of the   periodic structure with a period $ d=a+b $, where $ a $ and $ b $ are the thicknesses of the alternative layers of the periodic structure made of light and heavy material, respectively. In our case, the ratio of the thickness of the light layer $ a $ to the thickness of the 
  heavy layer $ b $ is the parameter $\beta = a/b = 2$ and $\chi_a(\hbar\omega)$, and $\chi_b(\hbar\omega)$ are their corresponding susceptibilities (Figure 1).   Further details  are described in Appendix ~\cite{nasonov5,nasonov6}.
 
    Consider the main spectral and angular properties of DTR and PXR presented in Figure 2. The patterns demonstrate the dependence of the energy of emitted photons (shown in color) and the shape of the angular distribution on the target orientation over a range of target inclination angles  from $\theta_0 = 30^o$ to $\theta_0 = 90^o$ (normal beam incidence on the target).  The calculations were performed in accordance with ~\cite{nasonov5,nasonov6}. The angular photon density and photon energy at the maximum of the angular distributionfor a given direction of emission ($\theta_{Dx}, \theta_{Dy}$) are shown by contour lines and color, respectively. The "X" sign is the multiplicity of an increase in the angular density scale. For example, the scale of density  radiation  for PXR is 10-fold less than that for DTR at $\theta_0 = 90^o$. The values of the maximum angular density for each pattern in Figure 2 are presented in Table 1.

\begin{figure}[htp]
    \centering
    \includegraphics[width=150mm]{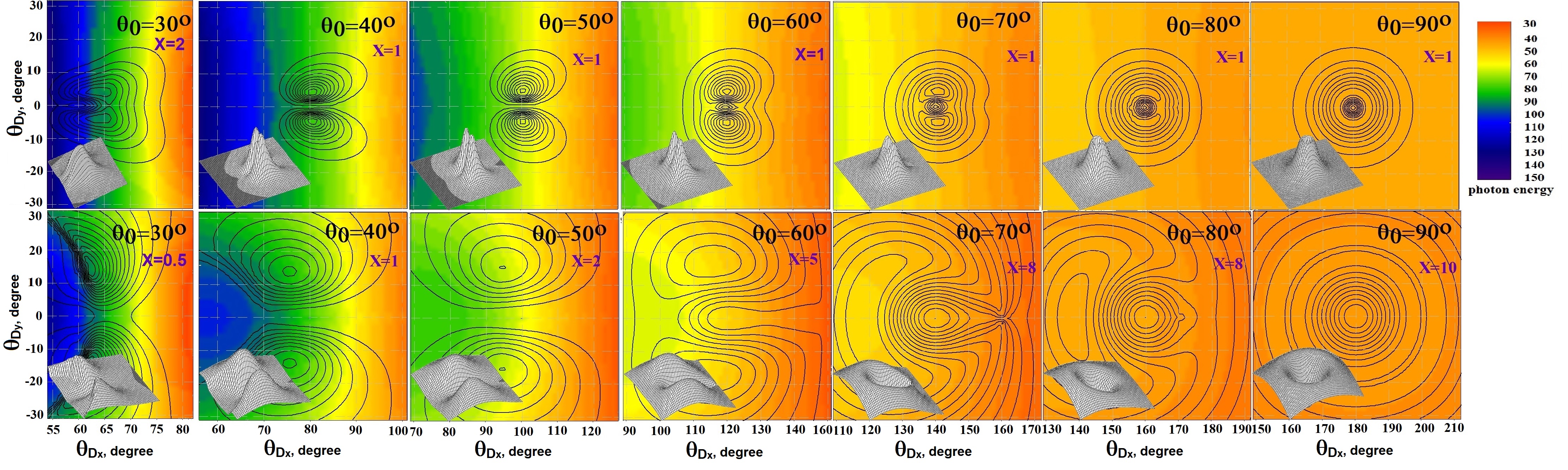}
    \caption{3D images of the full angular distribution of DTR and PXR generated at several tilt angles: from $\theta_0 = 30^o$ to normal beam incidence on target at $\theta_0 = 90^o$. 
    }  
    \label{fig:g2}
\end{figure}

\begin{table}[htbp]
\begin{center}
\caption{$dN/d\Omega$ DTR and PXR.\label{tab:i} }
\begin{tabular}{l|c lc lc lc lc lc lc lc lc}
\hline
$dN/d\Omega$, & & & & \underline{$\theta_0$, degree}\\
ph./sr
 &30 & \ \ \ \ 40 & 50 & \ \ \ \ 60 & 70 & \ \ \ \ 80 & 90 \\
\hline
DTR  & 0.381e-3 & 1.064e-3 & 1.126e-3 & 1.054e-3 & 0.937e-3 & 0.909e-3 & 0.849e-3\\
\hline
PXR  & 1.48e-3	& 0.916e-3 & 0.351e-3 & 0.171e-3 & 0.128e-3 & 0.126e-3 & 0.963e-4\\
\hline
\end{tabular}
\end{center}
\end{table}

    As can be seen from the figures, the annular crater-like shape of DTR (upper row of patterns) and PXR (lower row of patterns)  turns into a two-humped shape at decreased angle of inclination. For DTR, the emission maximum can be observed at the $d\theta_D=\gamma^{-1}$, while the PXR maximum depends on the energy of the generated photons and is observed close to $d\theta_{D}=(\gamma^{-2}  + |\chi^\prime_0(\hbar\omega)|)^{0.5}$ ~\cite{ivash, schag} around $\theta_D=2\theta_0$ ($\gamma$ is the Lorentz factor of electron). The maximum angular density radiation of DTR changes slightly with the target tilt angle. The angular density of PXR increases by about 15 times as the angle $\theta_0$ decreases from $90^o$ to $30^o$ (Table 1). For example, at $\theta_0 = 90^o$  the angular density of PXR is 10-fold less than that of DTR, whereas at $\theta_0 = 40^o$ the angular densities of PXR and DTR are almost equal. With further decrease in the angle $\theta_0 $, the density of PXR becomes approximately 4-fold greater than that of DTR.

The change in shape can be qualitatively explained by the dependence of the scattering coefficients of the $\pi$  and $\sigma $  components of the  line polarization of the radiation on the target inclination angle.
The $\pi $ component corresponds to the component of the electric field parallel to the diffraction plane $ {[\vec V \vec g]} $ (for a “symmetric” target $ [\vec V \vec g] $ || $[\vec V \vec N ]$, where $\vec g $ is the reciprocal vector of the periodic structure, $ \vec N $ is the normal to the target surface).
The contributions of the mutually perpendicular $\pi$ and $\sigma$ components  are determined by the factor $ C_p $, which are $C_\pi = cos^2(2\theta_0)$ and $ C_\sigma =1$, respectively.
The contrebution of the $ \pi $ component dominates in the $[\vec V \vec g] $ plane. The contribution of the $ \sigma $ component dominates in the plane, which is perpendicular to the $[\vec V \vec g] $ plane and passes through the axis of the radiation cone. 
Consequently, the factor $ C_p $ can affect the dynamics  of the angular distributions in the $[\vec V \vec g] $ plane at the changed angle $ \theta_0 $ . For this reason, the experimental study of the spectral and angular properties of radiation based on measurements carried out in the $[\vec V \vec g] $ plane is not so unambiguous. To simplify the interpretation of the experimental results by means of suppressing the dependence of radiation on the factor $C_p$, the angular distributions for different $\theta_0$ should be measured at a certain angle $ \theta_{Dy}$  to the $[\vec V \vec g] $ plane, for example, at $\theta_{Dy}=1/\gamma $. In this case, the detector scan path runs through the expected local maximum of the full angular distribution. This scan path is plotted in the 3D-image of the full angular distribution in Figure 3 by the black line. The intensity at the maximum of the angular distribution consists only of the $\sigma $ component with $C_p$=1. 

\begin{figure}[htp]
    \centering
    \includegraphics[width=85mm]{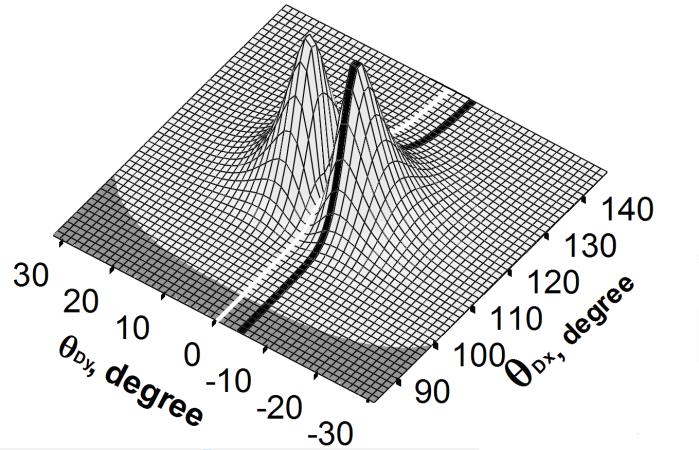}
    \caption{ 3D-image of the full angular distribution of RPS generated at $\theta_o = 57.5^o$.}  
    \label{fig:g3}
\end{figure}


 As can be seen from the figures, a changed tilt angle of the target changes the radiation spectrum, evidenced by both the changed radiation energy at the angular distribution maximum and the increased width of the emitted spectrum integrated over the entire radiation cone.

   The width of the PXR and DTR spectral lines varies within the radiation cone from $\Delta\hbar\omega\sim 3$~eV at $\theta_0= 90^o$  to $ \Delta\hbar\omega$  by several tens of electron volts at $\theta_0= 30^o$. The value of the photon energy in the center of the angular distribution also varies depending on the tilt angle over a wide range from 54 eV to 120 eV in accordance with the Bragg diffraction law. The border of the color change from green to blue in the patterns corresponds to the photon energy near the L absorption edge of radiation in Si at $ \hbar\omega= \hbar\omega_{LSi} \sim 100$ eV. For practical use and verification of the existing theoretical concepts, it is important to take into account the dependence of the properties of the observed radiation on spectral and angular features of PXR and DTR contributions and on their interference.
   
   \begin{figure}[htp]
    \centering
    \includegraphics[width=85mm]{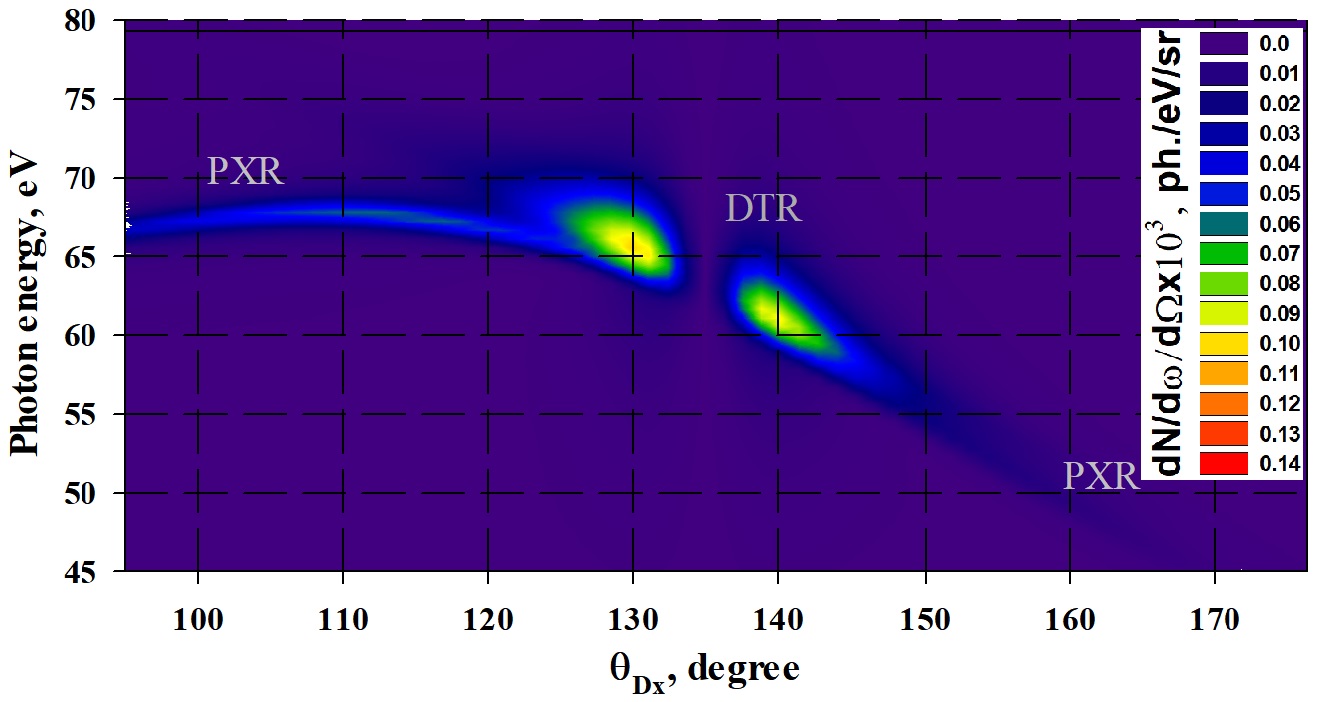}
    \caption{Spectral-angular dependence of the PXR and DTR output in the dffraction plane  $ [\vec V \vec g] $ at $ \theta_0=67.5^o $.}  
    \label{fig:g4}
\end{figure}

   Figure~4 illustrates the spectral and angular features of  PXR and DTR contributions when the angle between the electron beam and the target surface is $\theta_0=67.5^o$ . The figure shows the dependence of the PXR+DTR radiation intensity on the photon energy and the observation angle $\theta_{Dx}$ in the $[\vec V \vec g] $ plane. The results presented in Figure 4 show that the expected DTR contribution forms an angular distribution near its center, i.e., in the direction of mirror reflection of the electron velocity, while the PXR maxima are located on the periphery of the angular distribution. The maximum radiation intensity is observed to be  $\theta_{Dx}=2\theta_0 \pm 1/\gamma=135^o \pm 5.14^o$. In addition, it can be seen from the figure that the width of the PXR spectral line is narrower than that of the DTR spectral line.

\begin{figure}[htp]
    \centering
    \includegraphics[width=85mm]{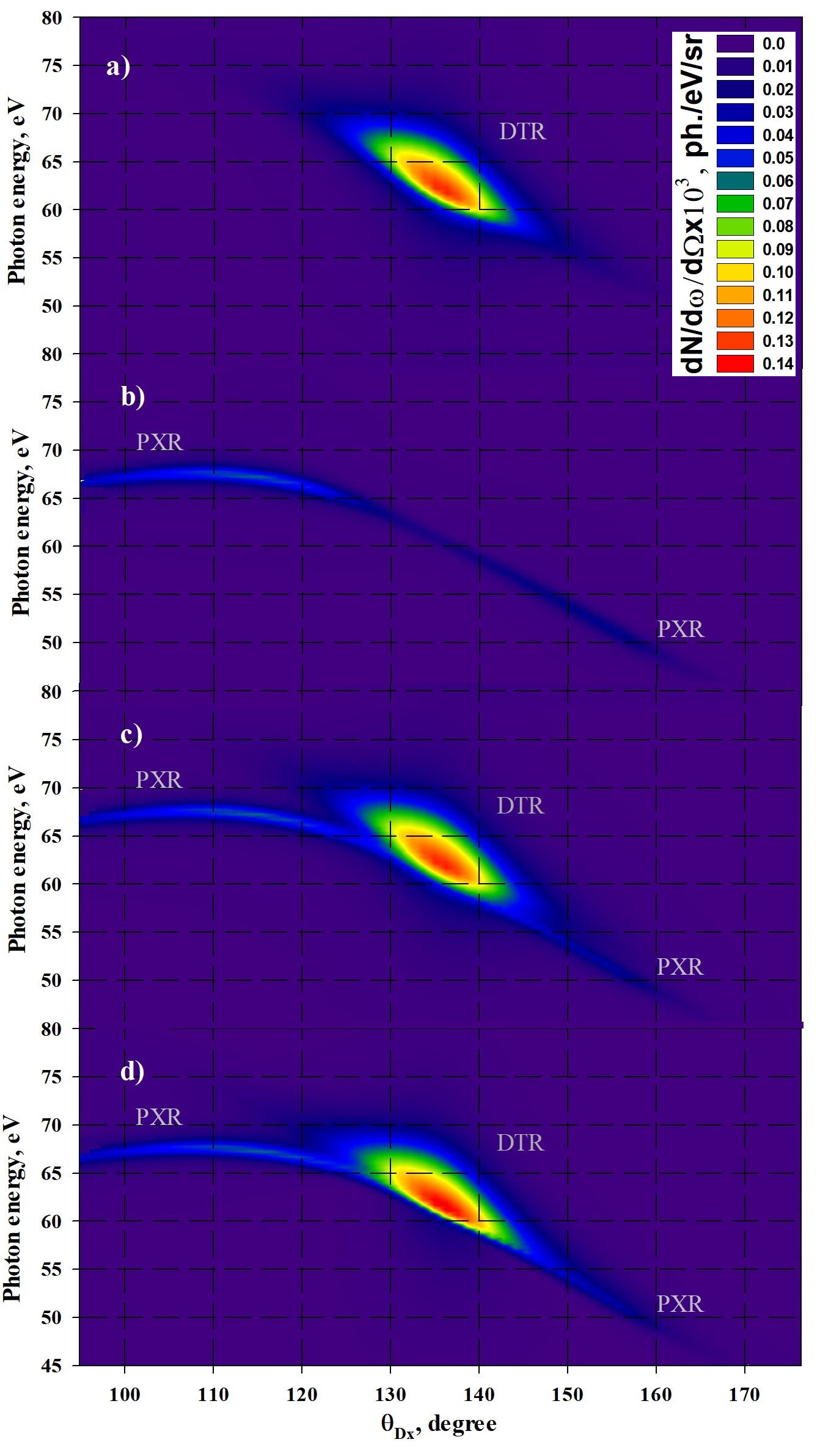}
    \caption{ Spectral-angular dependences of the radiation generated along $\theta_{Dx}$ for $\theta_{Dy}=1/\gamma$ and $\theta_0=67.5^o$: a) PXR; b) DTR; c) PXR+DTR; d) PXR+DTR with interference take into account.} 

    \label{fig:g5}
\end{figure}
   
    Figure 3 shows that the maximum radiation intensity can be  observed at $\theta_{Dx}=2\theta_0$ and $\theta_{Dy}= \pm1/\gamma= \pm 5.14^o$.  Consider the spectral and angular properties of DTR and PXR in the vicinity of the maximum radiation density.
     Figure 5 shows the spectral and angular properties of the contributions of  DTR and PXR  generated at $\theta_{Dy}= 1/\gamma$ to the $[\vec V \vec g]$ plane depending on the azimuth angle $\theta_{Dx}$. For comparison, Figure 5a and 5b separately show the contributions of PXR and DTR. Figure 5d and 5c show the contributions with and without interference, respectively. 
     As can be seen from the figures, the maximum intensity corresponds to the DTR contribution.
     The PXR contribution represents a long "whiskers" with a narrower spectral line width $\Delta E \sim 1.0 $~eV compared to the DTR spectral line width $\Delta E \sim 5.7 $~eV. The width of the PXR spectral line is smaller than that of DTR due a greater number of layers of the periodic structure involved in PXR generation compared to DTR generation. 
         The DTR+PXR pattern   with   regard to the interference is shown in Figure~5d. 
There are constructive and destructive interference on different parts on the spectral angular distribution of the PXR and DTR. 
    For example, from a comparison of Figures 5c and 5d, it is clear that the impact of interference somewhat  reduces the width of the spectral line, which is particularly apparent around $\theta_{Dx} =125^o$ and $\theta_{Dx} =145^o$.
                         
     In this paper, we did not consider the impact of multiple scattering of the electrons  on DTR and PXR. However, it should be noted that, unlike DTR, multiple scattering is expected to affect the properties of PXR.

\section{On the experimental observation of the fine structure of the spectrum  }

As can be seen from the figures, the maxima of the PXR and DTR contributions are separated by significant angular distances. The large angular distance between the PXR and DTR maxima makes it possible to confidently separate these contributions in the experiment and study their spectral properties. The measured spectral line width is an experimental evidence of the separately observed PXR and DTR contributions, since the width of the PXR spectral line should be several times narrower than that of the DTR (Figures 4 and 5).

\begin{figure}[htp]
    \centering
    \includegraphics[width=85mm]{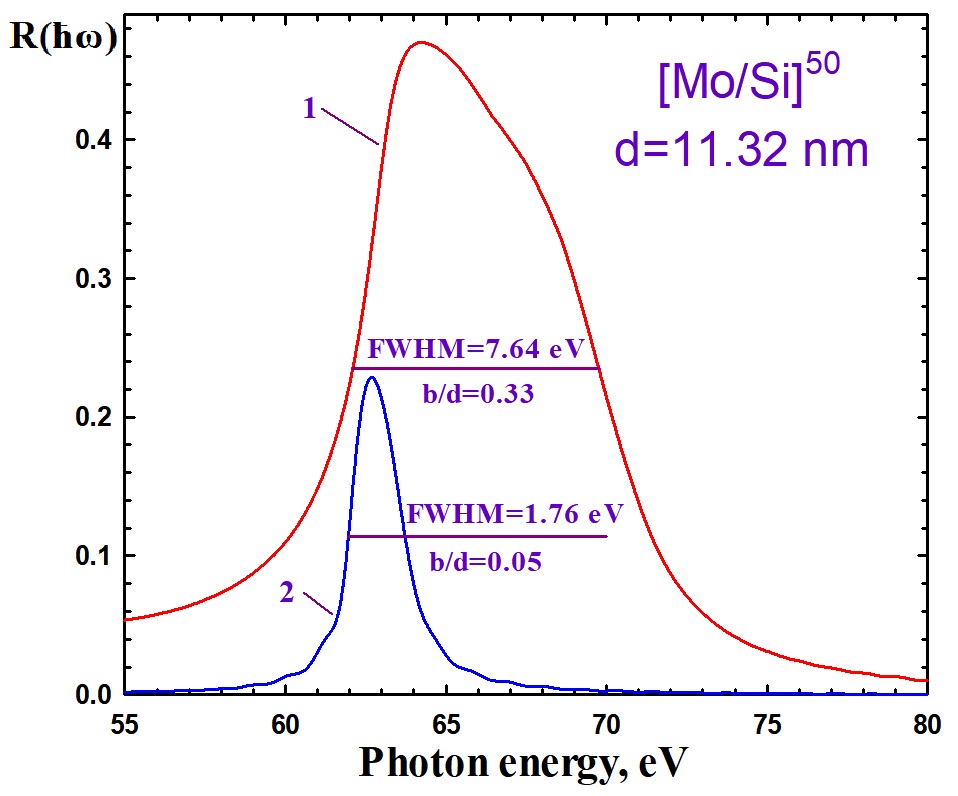}
    \caption{Reflectance curves $R(\hbar\omega)$ of the $[Mo/Si]^{50}$ X-ray mirror at $\theta_0=65^o$; curve 1- $b/d =0.33$ ; curve 2 - $b/d=0.05$. }  
    \label{fig:g6}
\end{figure}

The DTR and PXR spectra can be studied using another multilayer structure as a spectrometer~\cite{moran, knulst, AlCh,BeSiCh}. The width of the spectral line can be found experimentally by measuring the width of the rocking curve – the dependence of the radiation output on the tilt angle of the multilayer structure. The energy resolution of the spectrometer can be estimated from the calculation of the reflection coefficient $ R(\hbar\omega)$ depending on the photon energy for a given tilt angle of the structure. Let us estimate the energy resolution of the spectrometer based on the $ [Mo/Si]^{50} $ structure with a period $d$=11.3 nm, as an example. As is known ~\cite{Vino}, the resolution of this spectrometer depends on the ratio of the thickness of the heavy layer $b$ to the structure period $d$. Figure 6 shows the calculated dependences of the reflection coefficient $ R(\hbar\omega)$ of the $ [Mo/Si]^{50} $ X-ray mirror.  The calculations were performed using the method of recurrent relations ~\cite{Parrat,Vino, Kohn} for the tilt angle $\theta _0= 65^o$. Curve 1 was calculated for the mirror with $b/d $ = 1/3, similar to the structure used here as a target. Curve 2 was calculated for the structure with $ b/d $=0.05. The figure shows that FWHM of curve 1 is $\Delta E$= 7.64 eV and FWHM of curve 2 is $\Delta E$=1.76 eV at $ b/d $=0.3 and $b/d $=0.05, respectively. Hence, for the structure with $ b/d=0.05 $, the spectral line width may be sufficient to detect the difference between the spectral line widths of PXR and DTR.

It is necessary to indicate the contribution of characteristic VUV radiation from lines of target atoms Si and Mo. Data on the cross section for excitation of atomic levels by electrons ~\cite{epic} and the fluorescence yield ~\cite{xraylib} show that the angular fluorescence density of $Si_L$ and $Mo_M$ lines will be approximately $dN/d\Omega \sim 10^{-5}$ ph./sr. Thus, the radiation contributions from these lines are two orders of magnitude smaller than the expected contributions from DTR and PXR and can be ignored.

\section{Conclusion} 

The presented results show that the main contribution of the radiation generated in a multilayer mirror by electrons with an energy of several MeV comes from the DTR.  DTR has the size of the emission opening cone of about $\Delta \theta_{D} = \gamma^{-1}$ around the direction $\theta_{D}=2\theta_0$. At the same time, the maximum angular density of PXR is concentrated on the periphery of the total angular distribution of DTR+PXR with the size of the emission opening cone about $\Delta\theta_{D}=2(\gamma^{-2}  + |\chi^\prime_0(\hbar\omega)|)^{0.5}$.The results also show that the natural spectral line width of PXR is several times narrower than that of DTR.The difference in the width of the spectral lines can be used to experimentally identify the contributions of PXR and DTR.

\acknowledgments
{The work was supported by the Russian Science Foundation,
grant No. 23-22-00187.}

\begin{appendix}

\section{ Basic Formulas}
Basic formulas related to spectral and angular characteristics of radiation generated by electrons in layered periodic nanostructures – “X-ray mirrors” – were numerically studied using the theory ~\cite{nasonov5,nasonov6} developed by N. Nasonov.   The following formulas were used to calculate the spectral and angular intensity of PXR and DTR:  
\begin{equation}
\label{eq1}
\omega\frac{ {d N_\lambda(\omega, \theta)}}{{d\omega d\Omega}}  =  \left|A_\lambda\right|^{2} \ ,
 \end{equation}
 where
 \begin{align}
\label{eq2} 
A_\lambda = \frac{e}{\pi}\frac{\Omega_\lambda \alpha_\lambda}{(\tau+f_\lambda-i\delta)e^{i\eta f_\lambda}- (\tau-f_\lambda-i\delta)e^{-i\eta f_\lambda}} \times \nonumber \\ 
\times \begin{bmatrix}(\frac{1}{\gamma^{-2}+\Omega^2}- \frac{1}{\gamma^{-2}-\chi_0^\prime+\Omega^2})(e^{-i\eta(\sigma-\tau-f_\lambda)}-e^{-i\eta(\sigma-\tau+f_\lambda)})-\\ 
-(\frac{1}{\gamma^{-2}-\chi_0^\prime+\Omega^2}\frac{\tau+f_\lambda}{\sigma-\tau-f_\lambda}(1-e^{-i\eta(\sigma-\tau-f_\lambda)})-\\ 
-\frac{1}{\gamma^{-2}-\chi_0^\prime+\Omega^2}\frac{\tau-f_\lambda}{\sigma-\tau+f_\lambda}(1-e^{-i\eta(\sigma-\tau+f_\lambda)}))
\end{bmatrix}.
\end{align}
The first addend in square brackets corresponds to DTR, while the second addend corresponds to PXR.
Here, $ \hbar\omega$  and $ k_g = \hbar \omega/c\hbar$ are the energy and wave vector of the observed radiation, $\alpha_1= 1 $ , $\alpha_2=cos(2\theta_0) $  are the  polarization coefficients, $\theta_0$  is the angle between the electron velocity vector $ \vec V $ and the target surface, and $\vec N $ is the normal to the target surface; 
\begin{equation}
  \Omega_1= \frac{(\vec {e}_{1g} \vec{V}^\prime)}{\left|\vec{V}^\prime\right|}, \ \ \ \ \Omega_2= \frac{(\vec {e}_{2g} \vec{V}^\prime)}{\left|\vec{V}^\prime\right|},\ \ \ \vec{V}=\vec{V}_{||}+\vec{V}_{g}, \ \  \vec{V}^\prime=\vec{V}_{||}-\vec{V}_{g}, 
  \end{equation}  
  where $\vec V_{||}, \vec {V_g}$   are the  projections of the velocity vector on the target surface and on the direction of the reciprocal lattice vector $\vec g=2\pi/d $; $ d=a+b $ ; $ a $ and $ b $ are the thicknesses of alternating layers in the periodic structure – substances of low and high density, respectively; 
\begin{equation}
    \vec {e}_{1g}=\frac{[\vec{k}_g \vec{N}]}{\left|\vec{k}_g \right|},  \ \  
    \vec {e}_{2g}=\frac{[\vec{k}_g \vec {e}_{1g}]}{\left|\vec{k}_g \right|},
    \end{equation}
    \begin{equation}
    \eta=\frac{\left|\chi_g\right|(\hbar \omega)^2L}{g(c\hbar)^2}, \ \ 
    \delta=\frac{\chi_0^{\prime\prime}}{\left|\chi_g\right|}, \ \ 
    \sigma=\frac{1}{\left|\chi_g\right|}(\gamma^{-2} -\chi_0^\prime+\Omega^2),  \ \  
    \Omega^2=\Omega_1^2+\Omega_2^2,
    \end{equation}
here $\Omega_1$  and $\Omega_2$ are $\theta_{Dy}$ and $\Delta\theta_{Dx}= \theta_{Dx}- 2\theta_0$, respectively;
    \begin{equation}
        \tau=\frac{g^2(c\hbar)^2}{(\hbar\omega)^2\left|\chi_g\right|}(1-2\frac{\chi^\prime_0(\hbar\omega)^2}{g^2(c\hbar)^2}-\frac{\hbar\omega}{\hbar\omega^\prime_B}),
        \end{equation}
          \begin{equation}
        \hbar\omega_B^\prime=\hbar\omega_B(1+\Omega_{2})cot(\theta_0),
        \end{equation}
        here $\hbar\omega_B= gc\hbar/ 2sin(\theta_0)$ is the Bragg energy;
\begin{equation}
    f_\lambda=\sqrt{\tau^2-\alpha^2_\lambda-2i\delta(\tau-k_\lambda)}, \ \ 
    k_\lambda= \frac{sin(\pi\frac{a}{d})(\chi_a^{\prime\prime}- \chi_b^{\prime\prime})\alpha_\lambda^2}{\pi\chi_0^{\prime\prime}},
\end{equation}
  \begin{equation} 
      \chi_0=\frac{a}{d}\chi_a+\frac{b}{T}\chi_b, \ \  
      \chi_g=\frac{sin(m\pi\frac{a}{T})}{m\pi}(\chi_a-\chi_b),
  \end{equation} 
\begin{equation}
    \chi_{a,b}= \epsilon_{a,b}-1, \ \ \epsilon_{a,b}=n_{a,b}^2, \ \ 
    n_{a,b}=1-\frac{2\pi(c\hbar)^2\rho_{a,b}N_{av} r_e}{M_{a,b}(\hbar\omega)^2}(F_1+iF_2)_{a,b},
\end{equation}
where $ m $ is the diffraction order;$ F_1, F_2$ are the real and imaginary parts of the atomic scattering factor for photons.
\end{appendix}



\begin{thebibliography} {99}

\bibitem {uglov6!} S.R. Uglov, V.V. Kaplin, L.G. Sukhikh, A.V. Vukolov, { JETP Letters } {\bf 100}, No.8 503 (2014),DOI:10.7868/S0370274X14200053.          
\bibitem {uglov12} S.R. Uglov, V.V. Kaplin, A.P. Potylitsyn, L.G. Sukhikh, A.V. Vukolov and G. Kube, { Jour. of Phys.: Conf. Ser.} {\bf 517}, 012009  (2014), DOI:10.1088/1742-6596/517/1/012009.

\bibitem {uglov8andre} S R Uglov, V V Kaplin, A S Kubankin, J-M Andre, K Le Guen, Ph Jonnard, S de Rossi, E Meltchakov and F Delmotte, { Journal of Physics: Conference Series} {\bf 732} 012017 (2016), DOI:10.1088/1742-6596/732/1/012017.
 
\bibitem {loos10a} H. Loos, R. Akre, A. Brachmann, F.-J. Decker, Y. T. Ding,D. Dowell, P. Emma, J. C. Frisch, A. Gilevich, G. R. Hays,P. Hering, Z. Huang, R. H. Iverson, C. Limborg-Deprey, A.Miahnahri, S. Molloy, and H.-D. Nuhn, \emph{ Proceedings of the Thirtieth Free Electron Laser Conference, Gyeongjuouth Korea} (JACoW, Gyeongju, 2008), THBAU01, p.485.
 
\bibitem {wesch10b} S. Wesch, C. Behrens, B. Schmidt and P. Schmüser, \emph{Proceedings of the 31st International Free Electron LaserConference (FEL 09), Liverpool, UK} (STFC Daresbury Laboratory, Warrington, 2009), p.619.   
 
\bibitem {kube10} L. G. Sukhikh, G. Kube, S. Bajt, W. Lauth, Yu. A. Popov, and A. P. Potylitsyn, { Phys. Rev. ST Accel. Beams} {\bf 17}, 112805  (2014), DOI:10.1103/PhysRevSTAB.17.112805  

\bibitem {nasonov5}  N.N. Nasonov, V.V. Kaplin, S.R. Uglov, M.A. Piestrup, C.K. Gary,  {Phys. Rev. E} {\bf 68},036504  (2003), DOI:10.1103/PhysRevE.68.036504.
\bibitem {nasonov6}       N. Nasonov, V. Kaplin, S. Uglov, V. Zabaev, M. Piestrup, C. Gary, { Nucl. Instrum. and Meth. B} {\bf 227}, 41 (2005), DOI:10.1016/j.nimb.2004.06.020.

\bibitem {ivash}
I.D. Feranchuk and A.V. Ivashin, \emph{ Theoretical investigation of the parametric X-ray features, J.Physique, France } {\bf 46}  (1985) 1981, DOI:10.1051/jphys:0198500460110198100.

\bibitem {schag}
A.V. Shchagin, N.A. Khizhnyak \emph{ Differential properties of parametric
X-ray radiation from a thin crystal, Nuclear Instruments and Methods in
Physics Research B } {\bf 119}  (1996) 115, DOI:10.1016/0168-583X(96)00311-4.


\bibitem {esrf20} B.L. Henke and E.M. Gullikson and J.C. Davis, Atomic Data and Nuclear Data Tables {\bf 54},181 (1993), DOI:10.1006/adnd.1993.1013

 \bibitem {esrf23}  www.esrf.eu/Instrumentation/software/data-analysis/xop2.3.

 \bibitem {moran}
M. A. Moran, B. A. Dahling, M. A. Piestrup, B. L. Berman, and J. O. Kephart, \emph{ AIP Conference Proceedings} {\bf 147}, 34 (1986), DOI:10.1063/1.35961.

\bibitem {knulst}
 W. Knulst, van der M. J. Wiel, O. J. Luiten and J.Verhoeven, \emph {Observation of narrow-band Si L-edge 275 Cerenkov radiation generated by 5 MeV electrons, Appl. Phys. Lett.} {\bf  79}  (2001) 2999, DOI:10.1063/1.1415049.

 \bibitem {AlCh}
S. Uglov, A. Vukolov, V. Kaplin, L. Sukhikh and P. Karataev, \emph{ Observation of soft X-ray Cherenkov radiation in Al,  EPL (Europhysics Letters)} {\bf  118 } ( 2017) 34002, DOI:10.1209/0295-5075/118/34002.

\bibitem {BeSiCh}
S.R.Uglov, A.V.Vukolov, \emph{ Observation of soft X-ray Cherenkov radiation in Be and Si foil, JINST}{\bf 16} (2021) P07043 DOI:10.1088/1748-0221/16/07/P07043

\bibitem {Vino}
A.V. Vinogradov, I.V Kozhevnikov, \emph {Multilayer x-ray mirrors, Trudy FIAN} {\bf  196} (1989) 62.

\bibitem {Parrat}
L. G. Parratt, \emph{ Surface Studies of Solids by Total Reflection of X-Rays, Phys. Rev.} {\bf 95} (1954) 359, DOI:10.1103/PhysRev.95.359.
		
\bibitem {Kohn}
V.G. Kohn, \emph{ Towards the theory of specular reflection of X-rays by multilayer mirrors, Journal of  Surface Investigation. X-Ray, Synchrotron and Neutron Techniques} {\bf 1}  (2003) 23.


\bibitem {epic} 
{https://inis.iaea.org/search/searchsinglerecord.aspx?recordsFor=SingleRecord\&RN=39105838)}

\bibitem {xraylib} 
{http://ftp.esrf.fr/scisoft/xraylib/xraylib\_tables\_v2.3.pdf}

\end{thebibliography}
\end{document}